\documentclass{article}

\usepackage{arxiv}

\usepackage[utf8]{inputenc} 
\usepackage[T1]{fontenc}    
\usepackage{hyperref}       
\usepackage{url}            
\usepackage{booktabs}       
\usepackage{amsfonts}       
\usepackage{nicefrac}       
\usepackage{microtype}      
\usepackage{lipsum}
\usepackage{graphicx}
\graphicspath{ {./images/} }

\title{Assessment of AI-Generated Pediatric Rehabilitation SOAP-Note Quality}

\author{
 Solomon Amenyo \\
  David R. Cheriton School of Computer Science\\
  University of Waterloo\\
  Waterloo, Ontario \\
  \texttt{samenyo@uwaterloo.ca} \\
   \And
 Maura R. Grossman \\
  David R. Cheriton School of Computer Science\\
  University of Waterloo\\
  Waterloo, Ontario \\
  \texttt{maura.grossman@uwaterloo.ca} \\
  \And
 Daniel G. Brown \\
  David R. Cheriton School of Computer Science\\
  University of Waterloo\\
  Waterloo, Ontario \\
  \texttt{dan.brown@uwaterloo.ca} \\
  \And
 Brendan Wylie-Toal \\
  School of Environment, Enterprise and Development\\
  University of Waterloo\\
  Waterloo, Ontario \\
  \texttt{bwylieto@uwaterloo.ca} \\
}

\begin{document}
\maketitle
\begin{abstract}
This study explores the integration of artificial intelligence (AI) or large language models (LLMs) into pediatric rehabilitation clinical documentation, focusing on the generation of SOAP (Subjective, Objective, Assessment, Plan) notes, which are essential for patient care. Creating complex documentation is time-consuming in pediatric settings. We evaluate the effectiveness of two AI tools; Copilot, a commercial LLM, and KAUWbot, a fine-tuned LLM developed for KidsAbility Centre for Child Development (an Ontario pediatric rehabilitation facility), in simplifying and automating this process. We focus on two key questions: (i) How does the quality of AI-generated SOAP notes based on short clinician summaries compare to human-authored notes, and (ii) To what extent is human editing necessary for improving AI-generated SOAP notes? We found no evidence of prior work assessing the quality of AI-generated clinical notes in pediatric rehabilitation.

We used a sample of 432 SOAP notes, evenly divided among human-authored, Copilot-generated, and KAUWbot-generated notes. We employ a blind evaluation by experienced clinicians based on a custom rubric. Statistical analysis is conducted to assess the quality of the notes and the impact of human editing. The results suggest that AI tools such as KAUWbot and Copilot can generate SOAP notes with quality comparable to those authored by humans. We highlight the potential for combining AI with human expertise to enhance clinical documentation and offer insights for the future integration of AI into pediatric rehabilitation practice and other settings for the management of clinical conditions.
\end{abstract}


\section{Introduction}
\subsection{SOAP Notes in Pediatric Rehabilitation Care}

The importance of clinical documentation in healthcare, particularly in pediatric rehabilitation, cannot be overstated. SOAP notes, which stands for Subjective, Objective, Assessment, and Plan, provide a structured framework for clinicians to systematically record patient interactions and treatment plans \cite{morreale2024ota, gogineni2019designing}. These notes are not merely bureaucratic exercises; they are vital tools that inform clinical decision-making, ensure continuity of care, and facilitate effective communication among healthcare providers \cite{morreale2024ota}. In pediatric rehabilitation, where children's health status can change rapidly, accurate and timely documentation is even more important \cite{hofacer2019clinical}. Clinicians must navigate child development, varying treatment modalities, and precise tracking of progress and outcomes, all of which emphasize the critical nature of high-quality SOAP notes in this field \cite{hofacer2019clinical}.

Accurate documentation is essential not just for individual patient care but also for broader treatment planning, and quality assurance within healthcare systems \cite{demsash2023health}. Inaccuracies or omissions in clinical notes can lead to miscommunication, which may result in diagnostic errors and inappropriate treatment, potentially jeopardizing patient safety \cite{demsash2023health}. Unfortunately, the increasing burden of documentation on healthcare providers has been linked to clinician burnout, a pressing issue worsened by the ongoing shortage of primary care physicians and the growing demands placed on them \cite{budd2023burnout}. Effective clinical documentation practices are, therefore, vital not only for enhancing patient-care outcomes but also for alleviating some of the administrative pressures faced by healthcare professionals.

\subsection{The Role of AI in Healthcare Documentation}

The dawn of artificial intelligence (AI) technologies has introduced new paradigms in clinical documentation practices \cite{bajwa2021artificial}. AI-driven systems, particularly those using natural language processing (NLP) and machine learning algorithms, are being developed to automate the generation of clinical notes \cite{alowais2023revolutionizing}, including SOAP notes. Such technologies offer the promise of improving documentation efficiency, reducing the time clinicians spend on paperwork, and potentially increasing the quality of the notes produced. The introduction of automated systems like CliniKnote \cite{li2024improving} highlights the potential for AI to enhance the training and evaluation of models in clinical note generation, thereby simplifying workflows in healthcare settings. However, the integration of AI into clinical documentation also raises significant questions regarding the quality and reliability of AI-generated content.

Assessing the quality of AI-generated SOAP notes is of paramount importance. Despite the potential advantages, there are concerns regarding the accuracy, completeness, and adherence of these automated notes to clinical guidelines. Recent studies have shown the challenges associated with AI-generated documentation, revealing a range of errors, including high rates of omissions and inaccuracies \cite{zhai2024effects, dwivedi2023opinion}. For example, a study exploring the capabilities of ChatGPT-4 in generating SOAP notes found that the model produced an average of 23.6 errors per clinical case, with omissions constituting the majority of these errors \cite{kernberg2024using}. Such findings highlight the need for rigorous evaluation of AI-generated notes, particularly in sensitive areas such as pediatric rehabilitation, where the implications of documentation errors can be profound.

\section{Background}

In this study, Microsoft's Copilot tool, a commercially available AI system (chatbot), was used to generate SOAP notes for pediatric occupational therapy sessions. These notes were then reviewed and edited by occupational therapists (OTs) to ensure accuracy and completeness. Similarly, KAUWbot, a custom-built AI tool developed in collaboration between KidsAbility and the University of Waterloo, was also used to generate SOAP notes. This tool offered the unique opportunity to evaluate both unedited and edited versions of the SOAP notes, allowing the research team to assess the baseline quality of AI-generated notes and to measure the added value of human editing.

The central problem this study aims to address is to assess the quality and reliability of AI-generated SOAP notes in pediatric rehabilitation. While AI tools like Copilot and KAUWbot hold promise for reducing clinician workload \cite{dimaio2024creation} and improving efficiency, it is important to evaluate whether these tools produce SOAP notes that meet the rigorous quality standards required in healthcare documentation. We focus on two research questions: (i) How does the quality of AI-supported SOAP notes compare to that of human-authored SOAP notes in pediatric rehabilitation settings? and (ii) To what extent is human editing necessary for refining AI-generated SOAP note drafts, and how does it impact their overall quality? We explore the role AI can play in clinical documentation and the degree to which human oversight remains necessary to ensure the quality and accuracy of SOAP notes in pediatric rehabilitation. We contribute to the broader discussion on integrating AI into clinical workflows and offer practical recommendations for using AI in pediatric rehabilitation documentation. 

\subsection{Pediatric Occupational Therapy Clinics} 
Pediatric OT aims to help children and youth who experience challenges in activities of daily life due to physical, cognitive, or developmental disabilities or delays \cite{novak2019effectiveness}. Common case types include developmental delays, autism spectrum disorder, sensory processing disorders, cerebral palsy, and traumatic brain injury \cite{us2021intellectual}. Clinics serve children across a broad age range, from infants to adolescents, with interventions tailored to each developmental stage. Infants may receive early intervention for conditions like cerebral palsy, while older children and teens may work on skills needed for school tasks or independent living \cite{smythe2021early}. Sessions address fine and gross motor skills, sensory processing, cognitive and social development, and self-care skills, often through engaging, play-based activities \cite{smythe2021early}.

Therapists in OT clinics use a variety of approaches, such as strengthening exercises, sensory integration therapy, adaptive equipment training, and activities of daily living (ADL) training \cite{smythe2021early}. Sessions typically last 30-60 minutes and are scheduled approximately once per week, with the duration of treatment varying based on the child's progress and specific needs \cite{sugden2020evaluation}.

\subsection{KAUWbot for Clinical Documentation}
KAUWbot (the custom model in our study) represents a new development in pediatric occupational therapy documentation. Designed specifically for KidsAbility clinicians, this model automates the conversion of bullet point scratch notes into fully drafted SOAP notes, reducing documentation time while maintaining note quality \cite{dimaio2024creation}. KAUWbot leveraged historical SOAP notes and synthetically-generated scratch notes for training. The initial datasets did not include corresponding scratch notes; therefore, a synthetic dataset was generated using the Llama 2 70B Chat model. The generation process involved collaboration with clinicians from KidsAbility to ensure the synthetic scratch notes reflected real-world clinical documentation practices \cite{dimaio2024creation}.

The model's training involved domain-adaptive pre-training (DAPT) and fine-tuning techniques \cite{dimaio2024creation}. For DAPT, historical notes were filtered, tokenized, and anonymized to train the model on relevant domain language. Fine-tuning employed pairs of historical SOAP notes and synthetic scratch notes to adapt the model to its specific task, framed as a causal language modeling process \cite{dimaio2024creation}. To determine the most effective configuration, the evaluation process compared models trained with two distinct fine-tuning paradigms: LoRA (Low-Rank Adaptation) fine-tuning and full fine-tuning. LoRA fine-tuning, designed for computational efficiency, updated a smaller subset of parameters, whereas full fine-tuning updated all parameters, enabling more extensive adaptation to the specific task \cite{dimaio2024creation}. Comparative analysis of these paradigms demonstrated that while LoRA fine-tuning offered advantages in resource efficiency, full fine-tuning produced outputs of higher quality, with the Llama 3 8B Instruct model emerging as the superior option based on both subjective and objective metrics \cite{dimaio2024creation}.

\subsection{Copilot for Clinical Documentation}
In recent years, integrating AI-driven tools like Microsoft's Copilot in clinical settings emerged as a potential solution to streamline documentation and improve productivity in healthcare \cite{CoPH}. Occupational therapists are tasked with managing extensive documentation requirements \cite{schwellnus2020solution}. Hence, AI tools present a valuable opportunity to simplify the documentation process while simultaneously improving the quality of clinical notes.

Copilot (the commercial model in our study) uses NLP to generate text based on prompts, making it suitable for applications in healthcare documentation \cite{scharp2024natural}. For example, in a clinical setting, Copilot can assist in drafting notes by generating information from clinician inputs and generating a structured narrative that adheres to professional standards \cite{BoI}.

\subsection{Challenges in Pediatric Rehabilitation Documentation}

Children with disabilities often require a multidisciplinary treatment approach involving physicians, therapists, nurses, and social workers \cite{philip2022survey}. Children with disabilities also change in their physical and cognitive abilities over time, requiring regular documentation for their evolving needs and treatment plans \cite{haarbauer2017service}.

Pediatric rehabilitation is itself constantly evolving, with new technology, therapeutic approaches, and evidence-based practices \cite{michmizos2017pediatric}. Documentation systems must adapt to new interventions, outcome measures, and interdisciplinary collaborations \cite{seko2020impact}.

Electronic health records (EHRs) have significantly advanced documentation practices, but challenges persist for pediatric rehabilitation \cite{schwartz2024electronic}. EHRs are often designed for general medical settings \cite{koscielniak2022evaluating}. Moreover, the documentation burden imposed by EHRs, often perceived as a barrier to clinical care, can further intensify the challenges \cite{schwartz2024electronic}. EHR systems are not specifically designed for pediatric rehabilitation settings. Consequently, professionals in these environments often face significant challenges, spending months or even years adapting these tools to meet their unique needs and workflows \cite{schwartz2024electronic}.

\section{Methodology}
This section outlines the methodological approach used to assess the quality of SOAP notes in pediatric rehabilitation. It details the study population, data collection procedures, transformation of scratch notes into SOAP notes, anonymization and human editing processes, rubric criteria for evaluation, and the overall assessment process.

\subsection{Study Population}
432 human-authored and AI-generated pediatric rehabilitation SOAP notes were collected at KidsAbility Centre for Child Development. 

\begin{table}[h!]
\centering
\caption{The Four Soap Note Pools and their Descriptions}
\label{table:1}
\begin{tabular}{|p{2cm}|p{2cm}|p{8cm}| }
 \hline
 \multicolumn{3}{|c|}{\textbf{SOAP Note Pools}} \\
 \hline
 \textbf{Note Pools} & \textbf{Number} & \textbf{Description}\\
 \hline
 \textbf{K} & 108 & Human-authored SOAP notes \\
 \textbf{E} &  108 & Copilot-generated SOAP notes (edited by an OT) \\
 \textbf{U}  &  108 & KAUWbot-generated SOAP notes (unedited) \\
 \textbf{T} &  108 & KAUWbot-generated SOAP notes (edited by an OT) \\
\hline
\end{tabular}
\end{table}

\subsection{Data Collection}
Rehabilitation in pediatric occupational therapy is inherently subjective, as it focuses on identifying the functional goals each child wishes to achieve but with which they struggle \cite{wade2020rehabilitation}. Clinicians collaborate with children and their families to develop individualized treatment plans to improve specific areas of function. Approximately half of the clients at KidsAbility are treated through school-based programs, where occupational therapists work directly with children in their educational environment. The remaining clients, primarily children under the age of four, receive in-centre treatment, where families bring the child to the clinic for therapy.

KidsAbility collected a total of 432 SOAP notes from the clinic's EHR system, prepared between January 2023 and June 2024, to ensure a comprehensive and representative sample. This collection covered a balanced distribution across three key pools: human-authored notes, AI-drafted unedited notes, and AI-drafted edited notes by OTs, providing a holistic view of the clinical documentation. All collected SOAP notes were rigorously anonymized. Personal identifiers and any other information that could potentially reveal patient identities were removed before the data was analyzed.

\subsection{Scratch Notes to SOAP Notes}
Occupational therapists frequently create scratch notes in clinical documentation; they are brief, bullet-point, and concise entries that focus on the key components of the therapy session \cite{arora2017scratch}. These scratch notes are fast, informal documents that memorialize crucial observations, patient responses, and major actions or interventions that occur during a session. Given the fast-paced nature of clinical settings, particularly in pediatric rehabilitation, OTs rely on these notes to ensure that important details are noted without disrupting the flow of therapy \cite{novak2019effectiveness}.

\begin{figure}
    \centering
    \includegraphics[width=\linewidth]{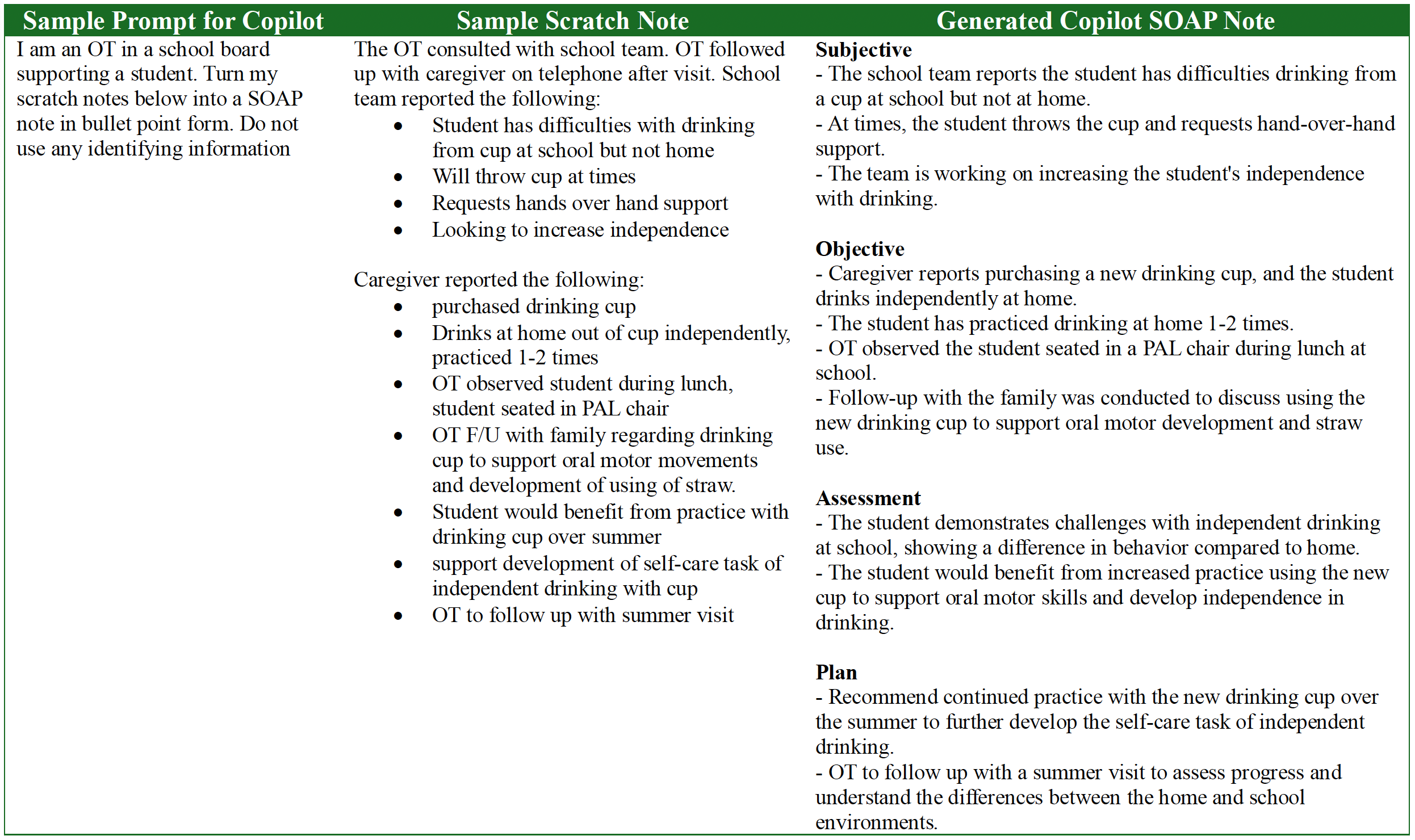}
    \caption{Sample Prompt for Copilot, Sample Scratch Note, and Generated Copilot SOAP Note.}
    \label{fig:1}
\end{figure}

After the therapy session is over, the OTs in our study enter their scratch notes into AI systems such as Copilot or KAUWbot, as shown in Figure \ref{fig:1}, to generate full SOAP notes. This tool analyzes and expands on the brief, shorthand entries provided by the OT, transforming them into detailed clinical documentation.

\subsection{SOAP Note Anonymization Process}

The anonymization of SOAP notes is necessary for maintaining the integrity of the evaluation process, ensuring unbiased assessments by the evaluators, and protecting client privacy. KidsAbility anonymized SOAP notes in three steps, each designed to conceal the specific origin of the notes while still allowing for systematic analysis.

\subsubsection{Step 1: Assigning Letters to Each Pool}
The first step in the anonymization process involved assigning specific letters to each pool of SOAP notes, ensuring that their origins remained undisclosed to evaluators. Human-authored notes were labeled "\textbf{K}," Copilot AI-generated notes edited by occupational therapists were labeled "\textbf{E}," unedited KAUWbot-generated notes were labeled "\textbf{U}," and edited KAUWbot notes were labeled "\textbf{T}."

\subsubsection{Step 2: Assigning Numbers to the Letters}
The notes were further masked by pairing the letter codes with random sequences of numbers. This process created unique alphanumeric codes, such as \textit{6N60}, \textit{N444}, or \textit{94N3}, ensuring that evaluators could not deduce a note's source from the letter alone.

\subsubsection{Step 3: Randomization of Notes}
Notes from all pools were thoroughly randomized, allowing evaluators to access any note from any pool at any time. This ensured that the evaluation process was not influenced by the sequence of notes or their grouping.

Through this anonymization process, our study achieved a robust level of confidentiality, ensuring that the evaluation of the SOAP notes was conducted in a fair and unbiased manner while maintaining patient privacy.

\subsection{Human Editing Process of SOAP Notes}
In the human editing process for both Copilot and KAUWbot-generated SOAP notes, OTs played a crucial role as the editors of the AI-produced content in ensuring the accuracy, relevance, and completeness. After the AI tools generated the initial draft of the SOAP notes based on the scratch notes provided, the OTs carefully reviewed the output to verify that the information was accurately reflected across all sections—Subjective, Objective, Assessment, and Plan. During this review, the OTs may have needed to make various corrections, such as fixing errors in the AI's interpretation of the input data or addressing inaccuracies in how patient observations or interventions were represented.

One common task during the editing process was the relocation of information from one section of the SOAP note to another. While AI tools like Copilot and KAUWbot are adept at processing and structuring data, they occasionally misplace details, assigning subjective observations to the objective section or assessment findings to the plan \cite{dwivedi2023opinion}. OTs needed to identify these errors and transfer the misplaced information to its appropriate section to ensure the logical flow and clinical accuracy of the documentation. 

In addition to correcting factual or structural errors, OTs also tailored the formatting and language of the AI-generated notes to align with their clinical style and preferences. Each clinician has a particular way of documenting information that may reflect institutional guidelines or personal preferences developed through years of practice. As such, the editing process may have involved reformatting the text, adjusting the tone, or adding specific terminologies that the AI may not have included. This personalization ensured that the final SOAP notes not only met clinical standards but also reflected the therapist's professional voice and documentation style.

\subsection{Rubric Criteria and Measurements}
The quality of SOAP notes was evaluated blindly based on five criteria, which were inspired by the PDQI-9 \cite{stetson2012assessing}; a shorter version of the original Physician Documentation Quality Instrument \cite{stetson2008preliminary}. The PDQI-9 measures documentation quality across nine key dimensions: \textit{completeness, correctness, appropriateness, organization, clarity, conciseness, comprehensiveness, usefulness,} and \textit{information synthesis}. These dimensions assess the overall quality of clinical notes to enhance their effectiveness in patient care. This study aimed to validate the PDQI-9 as a reliable and consistent tool for evaluating electronic clinical documentation and to highlight the importance of structured, accurate EHR entries in modern healthcare settings \cite{stetson2012assessing}.

Although the PDQI-9 provides a comprehensive framework for evaluating clinical note quality across nine dimensions, we focused only on five criteria (\textit{\textbf{Clear, Complete, Concise, Relevant,}} and \textbf{\textit{Organized}}) to evaluate the quality of SOAP notes, for several reasons. The simplicity of these criteria aligns more closely with the structured nature of SOAP notes, which emphasize simplified documentation for clinical efficiency. SOAP notes are meant to convey essential information succinctly and coherently \cite{notes2023updated}, and the chosen criteria directly reflect the core attributes needed for effective communication in this format. For example, clarity ensures that the note can be easily understood by other practitioners, while completeness and relevance ensure that all necessary and pertinent information is included without unnecessary details. Conciseness helps avoid redundancy and saves time, and organization ensures that information flows logically within the standardized SOAP format.
    
The evaluation instrument in this study featured a rubric comprised of five components, each rated on a three-point Likert scale. On this scale, a score of one indicated low quality, two signified moderate quality, and three represented the highest quality. The total score achievable ranged from a minimum of five to a maximum of 15.

\subsection{Evaluation Process}

Four independent evaluators, all experienced clinicians, assessed the quality of SOAP notes generated by both human-only and AI-assisted methods. To maintain objectivity and prevent any potential bias or influence, the evaluators were not allowed to communicate with each other throughout the evaluation process, and their identities were kept hidden from one another and from the OTs whose notes where analyzed. Each evaluator reviewed all of the SOAP notes, meaning that every SOAP note was evaluated a total of four times, once by each clinician. Given the total dataset of 432 SOAP notes from four different pools, this resulted in 1,728 individual evaluations.

Each SOAP note was accompanied by an evaluation instrument that incorporated the required criteria. This instrument provided evaluators with a standardized framework for assessment, ensuring that all notes were evaluated using the same rubric. The rubric covered essential aspects of the SOAP notes, such as clarity, completeness, conciseness, relevance, and organization. The rubric helped to promote fairness and accuracy across all assessments by providing a clear and consistent structure for evaluation.

\section{Results}
This section presents the results of SOAP note quality across different pools, beginning with descriptive statistics to summarize the distribution of quality scores. An ANOVA test is conducted to determine whether statistically significant differences exist between the quality scores of the various SOAP note pools.

\subsection{Descriptive Statistics for Quality Scores between SOAP Note Pools}

To evaluate the quality of SOAP notes in this study, we first examined certain descriptive statistics, which included the \textit{mean}, \textit{median}, and \textit{standard deviation} of scores by the four evaluators (Clinicians) for each group as shown in Table \ref{table:3}.

\begin{table}[h!]
\centering
\caption{Summary of Mean Values for Evaluators' (Clinician) Quality Ratings for the SOAP Note Pools.}
\label{table:2}
\begin{tabular}{|p{3cm}|p{2cm}|p{2cm}|p{2cm}|p{2cm}|}
 \hline
 \multicolumn{5}{|c|}{\textbf{SOAP Note Pool Mean Summary}} \\
 \hline
 \textbf{Evaluators} & \textbf{K}& \textbf{E} & \textbf{U}  & \textbf{T}\\
 \hline
 Clinician 1 & 11.80 & 12.57 & 12.97 & 13.21\\
Clinician 2 & 12.92 & 13.94 & 13.73 & \textbf{14.27}\\
Clinician 3 & \textbf{8.19} & 11.44 & 11.20 & 11.70\\
Clinician 4 & 12.06 & 13.38 & 12.73 & 13.32\\
 \hline
\end{tabular}
\end{table}

\begin{table}[h!]
\centering
\caption{Summary of Data for Evaluators' (Clinician) Quality Ratings for the SOAP Note Pools.}
\label{table:3}
\begin{tabular}{|p{2cm}|p{2cm}|p{2cm}|p{2cm}| }
 \hline
 \multicolumn{4}{|c|}{\textbf{SOAP Note Pool Data Summary}} \\
 \hline
 \textbf{Note Pools} & \textbf{Mean} & \textbf{Median} & \textbf{Std. Dev.}\\
 \hline
 \textbf{K} & 11.49 & 12.36 & \textbf{2.28}\\
 \textbf{E} &  12.83 & 12.98 & 1.09\\
 \textbf{U}  &  12.66 & 12.85 & 1.06\\
 \textbf{T} &  13.12 & 13.26 & 1.06\\
\hline
\end{tabular}
\end{table}

\begin{figure}
    \centering
    \includegraphics[width=\linewidth]{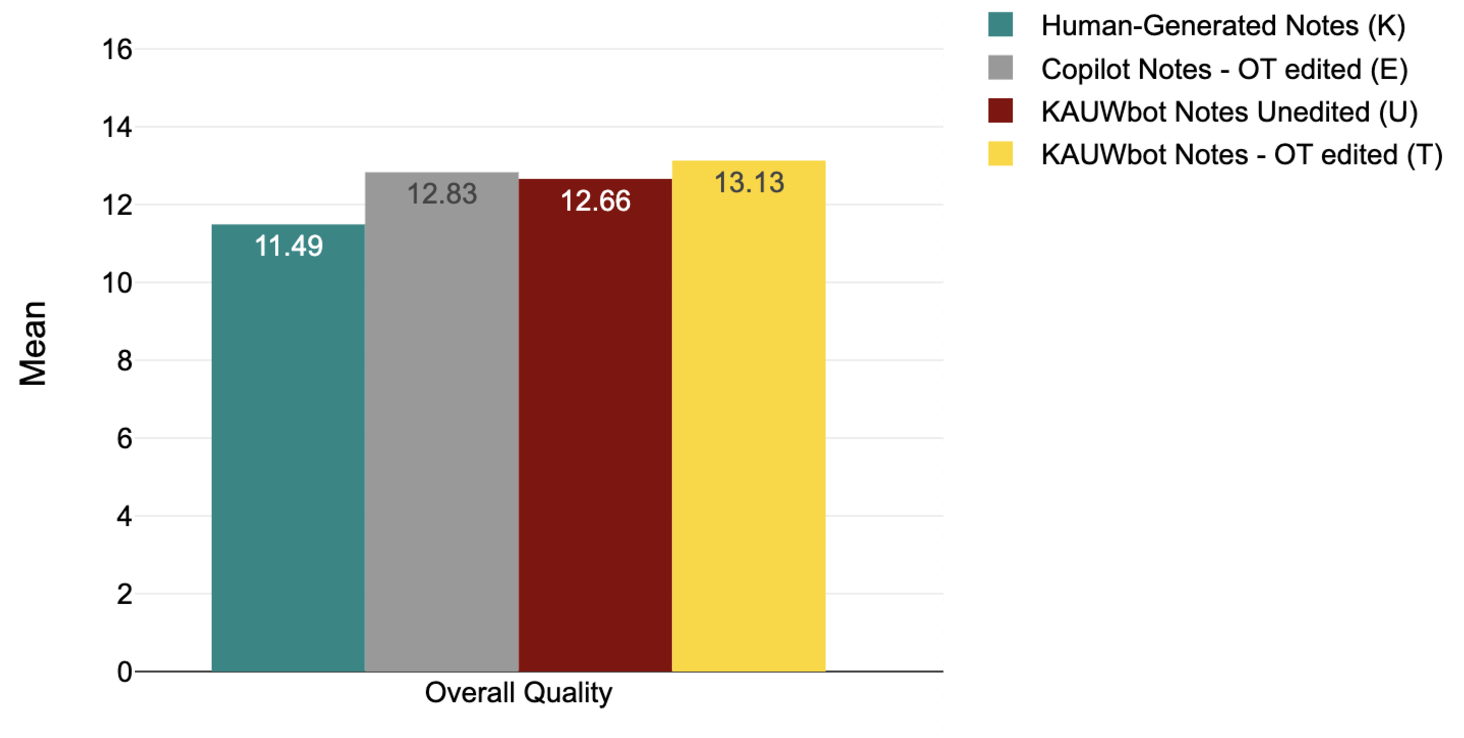}
    \caption{Histogram of Mean Quality Scores for the SOAP Note Pools.}
    \label{fig:2}
\end{figure}

\subsection{ANOVA to Compare the Quality Scores between SOAP Note Pools}

Analysis of Variance (ANOVA) is well-suited for this analysis because it enables a comparison of the quality scores across multiple groups—in this case, the four different pools of SOAP notes. ANOVA is ideal for this study as it allows simultaneous comparison of the means of more than two groups without inflating the risk of a Type I error (\textit{i.e., rejecting a true null hypothesis}) \cite{kao2008analysis}.

ANOVA is an effective method for assessing whether the differences in mean quality scores among the SOAP note pools are statistically significant or due to chance. It calculates an \textit{F-statistic} to determine if at least one group differs significantly. If the \textit{F-statistic} exceeds the critical value, it indicates the need for further post-hoc analysis to identify the specific groups with significant differences \cite{kao2008analysis}. 

A one-way ANOVA was conducted to assess the significance of the differences between the groups. The null hypothesis ($H_0$) in this analysis was that there was no significant difference in the mean quality scores across the different groups, while the alternative hypothesis ($H_1$) was that at least one group differed significantly.

\begin{table}[h!]
\centering
\caption{Summary of Analysis of Variance (ANOVA)}
\label{table:4}
\begin{tabular}{|p{3cm}|p{1cm}|p{1cm}|p{1cm}|p{1.5cm}|p{1.5cm}|p{1.5cm}| }
 \hline
 \multicolumn{7}{|c|}{\textbf{ANOVA Summary ($\alpha$ = 0.05)}} \\
 \hline
 \textbf{Source} & \textbf{DF} & \textbf{SS} & \textbf{MS} & $F-stat$ & $p-value$ & $F-crit$\\
 \hline
 \textbf{Between Groups} & 3 & 6.17 & 2.06 & 0.96 & 0.45 & 3.49\\
 \textbf{Within Groups} & 12 & 25.83 & 2.15 &  &  & \\
 \hline
 \textbf{Total}  & \textbf{15} & \textbf{32} &  &  &  &\\
\hline
\end{tabular}
\end{table}

There was no statistically significant difference in the mean quality scores between the four SOAP note pools. In other words, the differences observed in the mean scores ($K=11.49, E=12.83, U=12.66, T=13.12$) could be due to random variation rather than any actual effect of the different methods used to generate the SOAP notes. This suggests that, based on this analysis, the quality of SOAP notes generated by humans, Copilot (edited), KAUWbot (unedited), and KAUWbot (edited) were largely comparable, and no method significantly outperformed the others, though the overall mean score is highest for the KAUWbot (edited) pool.

\section{Discussion}
This section puts into perspective the study's findings by interpreting them in light of existing literature and their practical implications. It also discusses the study's limitations, potential directions for future research, and concluding insights on the quality of AI-assisted and human-authored SOAP notes in pediatric rehabilitation.

\subsection{Interpretation of Findings}
The descriptive statistics reveal trends in the quality scores across the four pools. KAUWbot-edited notes (T) had the highest mean score ($13.12$), followed by Copilot-edited notes ($E, 12.83$), unedited KAUWbot notes ($U, 12.66$), and human-authored notes ($K, 11.49$). This pattern suggests that incorporating AI tools, especially with subsequent human editing, may enhance the overall quality of SOAP notes and that the use of AI did not impair documentation quality in our study. The low variability (standard deviations ranging from 1.06 to 1.09) for AI-assisted notes compared to the higher variability in human-authored notes ($SD = 2.28$) indicates greater consistency in notes produced or refined with AI support.

Despite these apparent trends, the ANOVA results showed no statistically significant differences among the groups (\textit{F} = 0.96, \textit{p} = 0.45), indicating that the observed differences in mean scores could be attributed to random variation. 

\subsection{Implications of Findings}
\subsubsection{AI as a Complementary Tool in Clinical Documentation}
The results indicate the potential for AI tools like KAUWbot and Copilot systems to produce SOAP notes of comparable quality to human-authored notes, thereby increasing efficiency without reducing quality. When combined with human editing, these tools can achieve slightly higher average quality, as seen in the T and E pools. This suggests that AI can act as a valuable complement to human expertise, streamlining the documentation process without compromising quality.

\subsubsection{Consistency Across AI-Assisted Notes}
The lower variability in quality scores for AI-assisted notes highlights their ability to produce consistent outputs. This reliability could be particularly beneficial in busy clinical environments, where the quality of human-authored notes may vary due to fatigue, time constraints, or differing levels of expertise or attention.

\subsubsection{Human Oversight}
The higher scores for AI-generated notes edited by OTs compared to unedited versions (U) signify the importance of human oversight. Editing appears to address potential weaknesses in AI-generated documentation, ensuring alignment with clinical standards and specific requirements.

\subsubsection{Copilot vs. KAUWbot (Edited SOAP Notes)}
The mean scores of SOAP notes evaluated by the four clinicians highlight differences in the performance of the two AI systems. The scores suggest that KAUWbot, fine-tuned on pediatric OT data, produced slightly higher overall ratings than Copilot, a general-purpose language model. This indicates that fine-tuning an LLM on domain-specific data (pediatric OT) improves the quality of generated SOAP notes.

\subsection{Limitations and Future Directions}
One limitation of this study is the sample size, which, while substantial (432 SOAP notes), may not have provided sufficient statistical power to detect subtle differences in quality among the groups. With 108 notes per pool, the study may have been underpowered to identify required variations in mean scores. Increasing the sample size in future research could enhance the sensitivity of statistical analyses, allowing for more definitive conclusions about differences in quality across groups.

Another limitation of this research is the primary reliance on quantitative methods to assess the quality of SOAP notes, which may overlook the deeper insights that qualitative approaches could provide. While rubric-based evaluations and statistical analyses offer valuable metrics, a qualitative study could explore the subjective experiences and perceptions of OTs and evaluators regarding the usability, and practicality of AI-generated notes.

This study’s findings are specific to pediatric rehabilitation settings and thus may not be generalizable to other clinical contexts with different documentation requirements or evaluation criteria. Future research could extend this work to other specialties, exploring how AI tools adapt to various documentation styles and assessing the generalizability of the apparent benefit of human editing across fields.

\subsection{Conclusion}
This study provides compelling evidence that AI-assisted tools, particularly when augmented by human editing, can generate SOAP notes with quality levels comparable to those authored entirely by clinicians. The integration of AI tools, such as KAUWbot and Copilot systems, demonstrates their ability to simplify the documentation process while generating outputs that align with professional standards.

The findings show the potential for AI to alleviate some of the administrative burden on clinicians, allowing them to focus more on patient care without compromising the quality of their documentation. The consistency observed in AI-assisted outputs highlights their value in busy clinical environments, where time pressure and varying levels of clinician experience and attention can affect the uniformity of documentation. Overall, these results point to a future where AI tools may be easily integrated into clinical workflows, supporting healthcare professionals in maintaining high-quality documentation standards.

\section{Acknowledgements}
We extend our thanks to Rachel DiMaio, the developer of KAUWbot, for her support and dedication to this project. Her innovative contributions to the development of KAUWbot were instrumental in advancing this research.

Special thanks go to Ilona Manda, whose meticulous efforts in sorting the various SOAP notes, managing the anonymization process, and coordinating the evaluation workflow ensured the smooth execution of this study.

We acknowledge the four clinicians who contributed time and expertise to evaluate the quality of the SOAP notes. 

Finally, we are profoundly grateful to the KidsAbility Centre for Child Development for their invaluable support and collaboration. Their provision of resources and commitment to improving pediatric rehabilitation care were essential in making this research possible.

\clearpage
\bibliographystyle{unsrt}  
\bibliography{references}  

\begin{thebibliography}{10}

\bibitem{morreale2024ota}
Marie Morreale.
\newblock {\em {The OTA’s Guide to Documentation: Writing SOAP Notes}}.
\newblock Routledge, 2024.

\bibitem{gogineni2019designing}
Hyma Gogineni, Josephine~P Aranda, and Linda~S Garavalia.
\newblock {Designing Professional Program Instruction to Align with Students’ Cognitive Processing}.
\newblock {\em Currents in Pharmacy Teaching and Learning}, 11(2):160--165, 2019.

\bibitem{hofacer2019clinical}
Rylon~D Hofacer, Andrew Panatopoulos, Jared Vineyard, Rick Tivis, Elaine Nguyen, Niu Jingjing, and Ryan~P Lindsay.
\newblock {Clinical Care Coordination in Medically Complex Pediatric Cases: Results from the National Survey of Children with Special Health Care Needs}.
\newblock {\em Global Pediatric Health}, 6:2333794X19847911, 2019.

\bibitem{demsash2023health}
Addisalem~Workie Demsash, Sisay~Yitayih Kassie, Abiy~Tasew Dubale, Alex~Ayenew Chereka, Habtamu~Setegn Ngusie, Mekonnen~Kenate Hunde, Milkias~Dugassa Emanu, Adamu~Ambachew Shibabaw, and Agmasie~Damtew Walle.
\newblock {Health Professionals’ Routine Practice Documentation and its Associated Factors in a Resource-limited Setting: A Cross-Sectional Study}.
\newblock {\em BMJ Health \& Care Informatics}, 30(1), 2023.

\bibitem{budd2023burnout}
Jeffrey Budd.
\newblock {Burnout Related to Electronic Health Record Use in Primary Care}.
\newblock {\em Journal of Primary Care \& Community Health}, 14:21501319231166921, 2023.

\bibitem{bajwa2021artificial}
Junaid Bajwa, Usman Munir, Aditya Nori, and Bryan Williams.
\newblock {Artificial Intelligence in Healthcare: Transforming the Practice of Medicine}.
\newblock {\em Future Healthcare Journal}, 8(2):e188--e194, 2021.

\bibitem{alowais2023revolutionizing}
Shuroug~A Alowais, Sahar~S Alghamdi, Nada Alsuhebany, Tariq Alqahtani, Abdulrahman~I Alshaya, Sumaya~N Almohareb, Atheer Aldairem, Mohammed Alrashed, Khalid Bin~Saleh, Hisham~A Badreldin, et~al.
\newblock {Revolutionizing Healthcare: The Role of Artificial Intelligence in Clinical Practice}.
\newblock {\em BMC Medical Education}, 23(1):689, 2023.

\bibitem{li2024improving}
Yizhan Li, Sifan Wu, Christopher Smith, Thomas Lo, and Bang Liu.
\newblock {Improving Clinical Note Generation from Complex Doctor-Patient Conversation}.
\newblock {\em arXiv preprint arXiv:2408.14568}, 2024.

\bibitem{zhai2024effects}
Chunpeng Zhai, Santoso Wibowo, and Lily~D Li.
\newblock {The Effects of Over-Reliance on AI Dialogue Systems on Students' Cognitive Abilities: A Systematic Review}.
\newblock {\em Smart Learning Environments}, 11(1):28, 2024.

\bibitem{dwivedi2023opinion}
Yogesh~K Dwivedi, Nir Kshetri, Laurie Hughes, Emma~Louise Slade, Anand Jeyaraj, Arpan~Kumar Kar, Abdullah~M Baabdullah, Alex Koohang, Vishnupriya Raghavan, Manju Ahuja, et~al.
\newblock {Opinion Paper: “So What if ChatGPT Wrote It?” Multidisciplinary Perspectives on Opportunities, Challenges, and Implications of Generative Conversational AI for Research, Practice and Policy}.
\newblock {\em International Journal of Information Management}, 71, 2023.

\bibitem{kernberg2024using}
Annessa Kernberg, Jeffrey~A Gold, and Vishnu Mohan.
\newblock {Using ChatGPT-4 to Create Structured Medical Notes From Audio Recordings of Physician-Patient Encounters: Comparative Study}.
\newblock {\em Journal of Medical Internet Research}, 26:e54419, 2024.

\bibitem{dimaio2024creation}
Rachel DiMaio.
\newblock {Creation of a Custom Language Model for Pediatric Occupational Therapy Documentation}.
\newblock {Master's Thesis}, University of Waterloo, 2024.

\bibitem{novak2019effectiveness}
Iona Novak and Ingrid Honan.
\newblock {Effectiveness of Paediatric Occupational Therapy for Children with Disabilities: A Systematic Review}.
\newblock {\em Australian Occupational Therapy Journal}, 66(3):258--273, 2019.

\bibitem{us2021intellectual}
US~Department of~Health, Human Services, et~al.
\newblock {About Intellectual and Developmental Disabilities (IDDs)}.
\newblock {\em Eunice Kennedy Shriver National Institute of Child Health and Human Development (NICHD)}, 2021.

\bibitem{smythe2021early}
Tracey Smythe, Maria Zuurmond, Cally~J Tann, Melissa Gladstone, and Hannah Kuper.
\newblock {Early Intervention for Children with Developmental Disabilities in Low and Middle-Income Countries--The Case for Action}.
\newblock {\em International Health}, 13(3):222--231, 2021.

\bibitem{sugden2020evaluation}
Eleanor Sugden, Elise Baker, A~Lynn Williams, Natalie Munro, and Carol~M Trivette.
\newblock {Evaluation of Parent- and Speech-Language Pathologist--Delivered Multiple Oppositions Intervention for Children with Phonological Impairment: A Multiple-Baseline Design Study}.
\newblock {\em American Journal of Speech-Language Pathology}, 29(1):111--126, 2020.

\bibitem{CoPH}
Anna-Marie Brittain.
\newblock {Microsoft Copilot and the Future of Healthcare: AI-Driven Innovation}.
\newblock \url{https://www.hp.com/us-en/shop/tech-takes/microsoft-copilot-healthcare-future}, 2024.
\newblock [Online; accessed 30-October-2024].

\bibitem{schwellnus2020solution}
Heidi Schwellnus, Yukari Seko, Gillian King, Patricia Baldwin, and Michelle Servais.
\newblock {Solution-Focused Coaching in Pediatric Rehabilitation: Perceived Therapist Impact}.
\newblock {\em Physical \& Occupational Therapy in Pediatrics}, 40(3):263--278, 2020.

\bibitem{scharp2024natural}
Danielle Scharp, Mollie Hobensack, Anahita Davoudi, and Maxim Topaz.
\newblock {Natural Language Processing Applied to Clinical Documentation in Post-Acute Care Settings: A Scoping Review}.
\newblock {\em Journal of the American Medical Directors Association}, 25(1):69--83, 2024.

\bibitem{BoI}
Board of~Innovation.
\newblock {Microsoft Copilot}.
\newblock \url{https://healthcare.boardofinnovation.com/microsoft-copilot/}, 2024.
\newblock [Online; accessed 12-November-2024].

\bibitem{philip2022survey}
Kemly Philip and Glendaliz Bosques.
\newblock {A Survey of Inpatient Pediatric Rehabilitation Practices Across the United States 1}.
\newblock {\em Journal of Pediatric Rehabilitation Medicine}, 15(3):425--431, 2022.

\bibitem{haarbauer2017service}
Juliet Haarbauer-Krupa, Angela Ciccia, Jonathan Dodd, Deborah Ettel, Brad Kurowski, Angela Lumba-Brown, and Stacy Suskauer.
\newblock {Service Delivery in the Healthcare and Educational Systems for Children Following Traumatic Brain Injury: Gaps in Care}.
\newblock {\em The Journal of Head Trauma Rehabilitation}, 32(6):367--377, 2017.

\bibitem{michmizos2017pediatric}
Konstantinos~P Michmizos and Hermano~Igo Krebs.
\newblock {Pediatric Robotic Rehabilitation: Current Knowledge and Future Trends in Treating Children with Sensorimotor Impairments}.
\newblock {\em NeuroRehabilitation}, 41(1):69--76, 2017.

\bibitem{seko2020impact}
Yukari Seko, Gillian King, Sarah Keenan, Joanne Maxwell, Anna Oh, and CJ~Curran.
\newblock {Impact of Solution-Focused Coaching Training on Pediatric Rehabilitation Specialists: A Longitudinal Evaluation Study}.
\newblock {\em Journal of Interprofessional Care}, 34(4):481--492, 2020.

\bibitem{schwartz2024electronic}
Jessica Schwartz-Dillard, Travis Ng, Joann Villegas, Derrick Johnson, and Mary Murray-Weir.
\newblock {Electronic Documentation Burden among Outpatient Rehabilitation Therapists: A Qualitative Descriptive Study and Quality Improvement Initiative}.
\newblock {\em Journal of the American Medical Informatics Association}, 31(10):2347--2355, 2024.

\bibitem{koscielniak2022evaluating}
Nikolas~J Koscielniak, Carole~A Tucker, Andrew Grogan-Kaylor, Charles~P Friedman, Rachel Richesson, Josh~S Tucker, and Gretchen~A Piatt.
\newblock {Evaluating Completeness of Discrete Data on Physical Functioning for Children with Cerebral Palsy in a Pediatric Rehabilitation Learning Health System}.
\newblock {\em Physical Therapy}, 102(1), 2022.

\bibitem{wade2020rehabilitation}
Derick~T Wade.
\newblock {What is Rehabilitation? An Empirical Investigation Leading to an Evidence-Based Description}.
\newblock {\em Clinical Rehabilitation}, 34(5):571--583, 2020.

\bibitem{arora2017scratch}
Aarti~B Arora and Rebecca~L Bubp.
\newblock {Scratch Notes}.
\newblock {\em The International Encyclopedia of Communication Research Methods}, 2017.

\bibitem{stetson2012assessing}
Peter~D Stetson, Suzanne Bakken, Jesse~O Wrenn, and Eugenia~L Siegler.
\newblock {Assessing Electronic Note Quality Using the Physician Documentation Quality Instrument (PDQI-9)}.
\newblock {\em Applied Clinical Informatics}, 3(02):164--174, 2012.

\bibitem{stetson2008preliminary}
Peter~D Stetson, Frances~P Morrison, Suzanne Bakken, Stephen~B Johnson, and eNote Research~Team.
\newblock {Preliminary Development of the Physician Documentation Quality Instrument}.
\newblock {\em Journal of the American Medical Informatics Association}, 15(4):534--541, 2008.

\bibitem{notes2023updated}
Podder V, Lew V, and Ghassemzadeh S.
\newblock {SOAP Notes}.
\newblock {\em In: {StatPearls}}, 2023.

\bibitem{kao2008analysis}
Lillian~S Kao and Charles~E Green.
\newblock {Analysis of Variance: Is there a Difference in Means and What Does it Mean?}
\newblock {\em Journal of Surgical Research}, 144(1):158--170, 2008.

\end{thebibliography}

\clearpage
\appendix
\section{Appendix: Comparison of Evaluator's (clinician) Rubric Mean Scores for SOAP Note Pools}

The appendix includes detailed graphs illustrating the scoring patterns of each evaluator (\textit{clinician}) for the four SOAP note pools (\textit{Human-authored}, \textit{Copilot-edited}, \textit{KAUWbot-unedited}, and \textit{KAUWbot-edited}). These graphs visually represent the evaluations conducted using the custom rubric, which assessed the notes based on five key criteria: \textit{clarity}, \textit{completeness}, \textit{conciseness}, \textit{relevance}, and \textit{organization}. Each graph highlights the individual scoring trends, providing a detailed understanding of evaluator variability and offering insights into how different pools performed across the rubric’s dimensions.

\begin{figure}[h]
\centering
    \includegraphics[width=12cm]{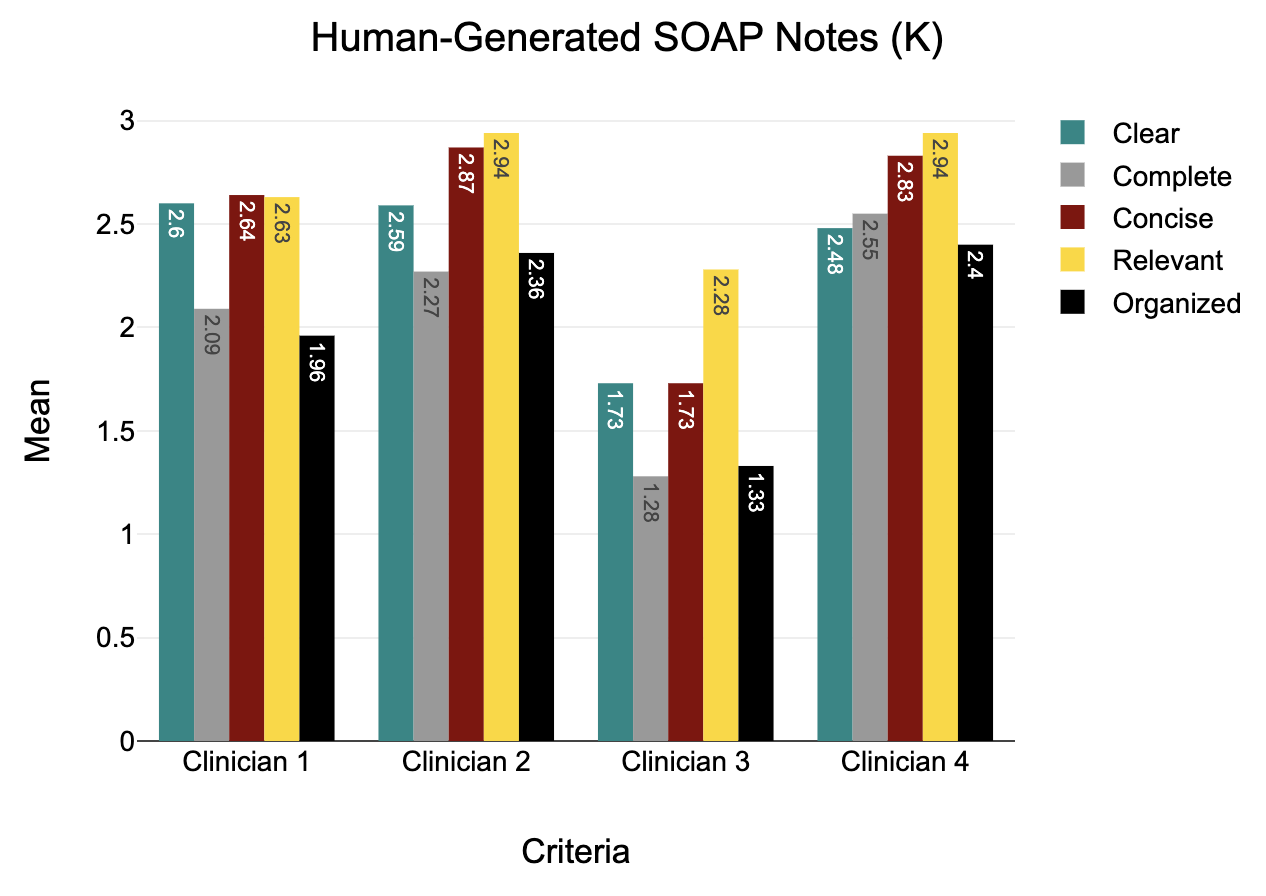}
\end{figure}

\begin{figure}[h]
\centering
    \includegraphics[width=12cm]{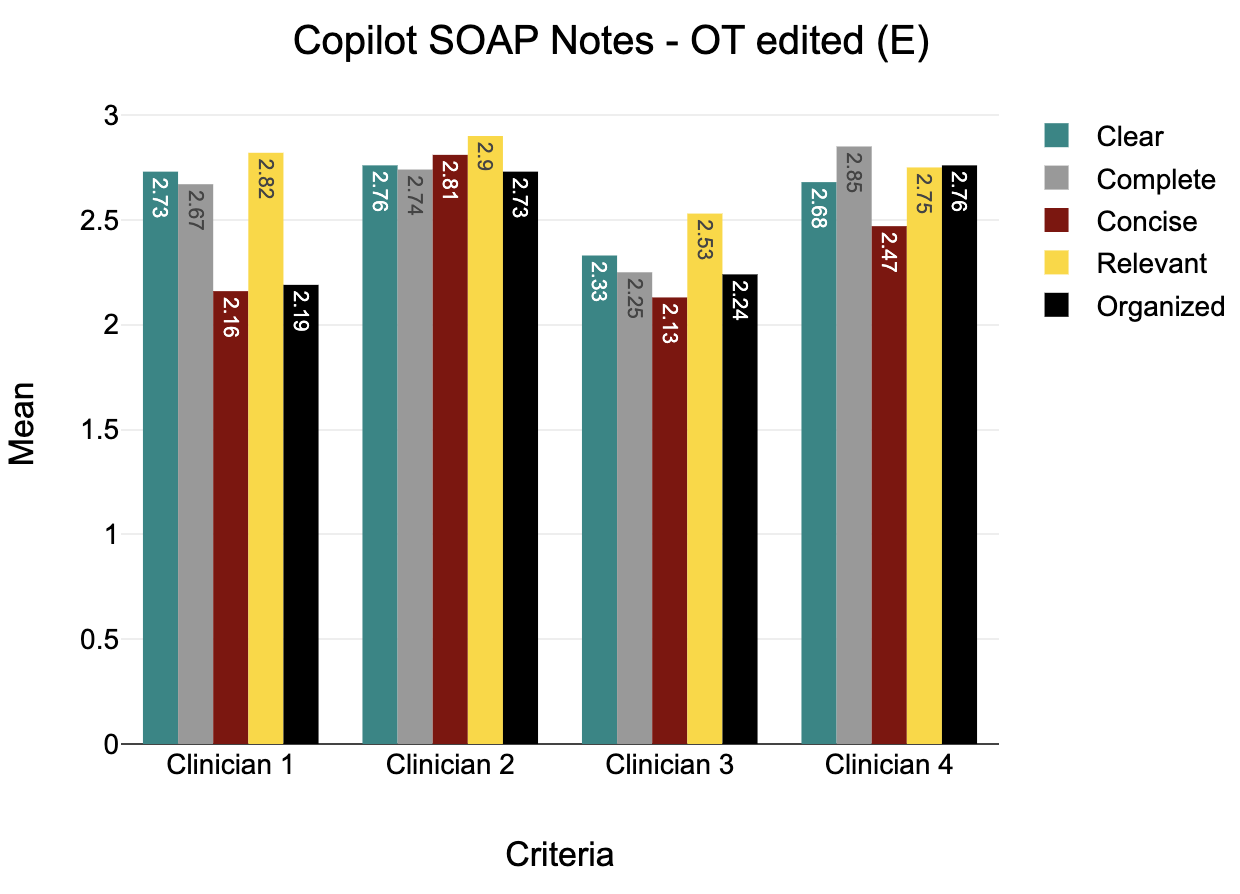}
\end{figure}

\begin{figure}[h]
\centering
    \includegraphics[width=12cm]{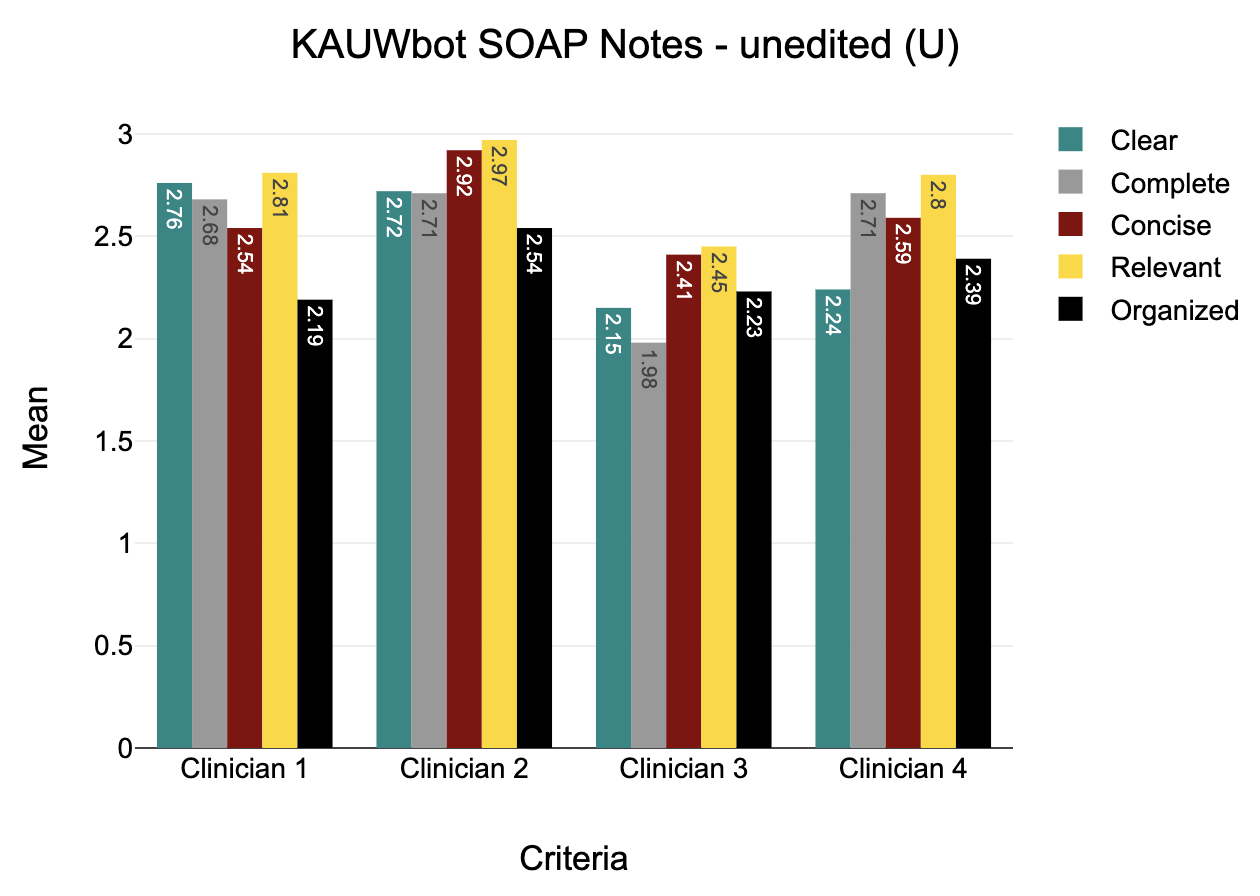}
     \vspace{1cm}
\end{figure}

\begin{figure}[h]
\centering
    \includegraphics[width=12cm]{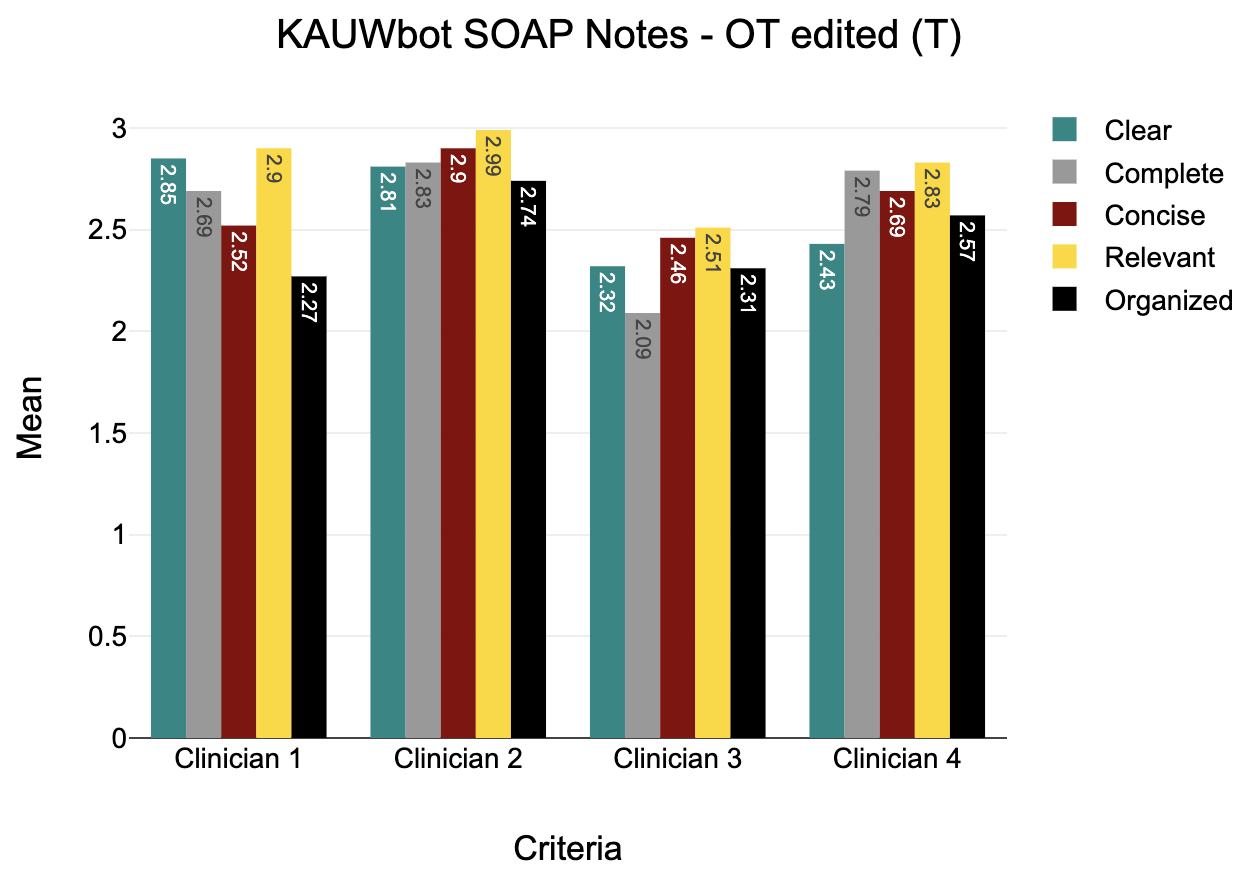}
\end{figure}

\end{document}